# DART Mission Determination of Momentum Transfer: Model of Ejecta Plume Observations

Andrew F. Cheng[1], Angela M. Stickle[1], Eugene G. Fahnestock[2], Elisabetta Dotto[3], Vincenzo Della Corte[4], Nancy L. Chabot[1], Andy S. Rivkin[1]

[1]JHU/APL, MD USA (andrew.cheng@jhuapl.edu), [2]JPL, CA USA, [3]INAF-OA, Rome, Italy, [4]INAF-IAPS, Rome, Italy

**ABSTRACT**

The NASA Double Asteroid Redirection Test (DART) spacecraft will impact the secondary member of the [65803] Didymos binary in order to perform the first demonstration of asteroid deflection by kinetic impact. Determination of the momentum transfer to the target body from the kinetic impact is a primary planetary defense objective, using ground-based telescopic observations of the orbital period change of Didymos and imaging of the DART impact ejecta plume by the LICIACube cubesat, along with modeling and simulation of the DART impact. LICIACube, contributed by the Italian Space Agency, will perform a flyby of Didymos a few minutes after the DART impact, to resolve the ejecta plume spatial structure and to study the temporal evolution. LICIACube ejecta plume images will help determine the vector momentum transfer from the DART impact, by determining or constraining the direction and the magnitude of the momentum carried by ejecta. A model is developed for the impact ejecta plume optical depth, using a point source scaling model of the DART impact. The model is applied to expected LICIACube plume images and shows how plume images enable characterization of the ejecta mass versus velocity distribution. The ejecta plume structure, as it evolves over time, is determined by the amount of ejecta that has reached a given altitude at a given time. The evolution of the plume optical depth profiles determined from LICIACube images can distinguish between strength-controlled and gravity-controlled impacts, by distinguishing the respective mass versus velocity distributions. LICIACube plume images discriminate the differences in plume structure and evolution that result from different target physical properties, mainly strength and porosity, thereby allowing inference of these properties to improve the determination of momentum transfer.

# 1. Introduction

The Double Asteroid Redirection Test (DART) mission will target the secondary member of the [65803] Didymos binary asteroid system in September - October 2022 in order to demonstrate asteroid deflection by kinetic impact, that is, modification of the orbit of the moon through momentum transfer. The DART kinetic impact on the 160-m secondary of the Didymos system will be the first hypervelocity impact experiment on an asteroid at a realistic scale relevant to planetary defense (Cheng et al. 2016, 2018). DART team members are part of the Asteroid Impact & Deflection Assessment (AIDA) international cooperation along with members of the ESA Hera mission, which will rendezvous with Didymos about 4 years after the DART impact to make detailed measurements of the Didymos system, including the mass of the secondary and the morphometry of the DART impact crater (Michel et al. 2016, 2018). The Deep Impact mission previously impacted Comet 9/PTempel 1 (A'Hearn et al. 2005), and the LCROSS mission (Colaprete et al. 2010, Schultz et al. 2010) hit the Moon, but neither mission caused, or was intended to cause, an observable orbit deflection, and these mission targets are much larger than any Potentially Hazardous Asteroid.

Significant updates to the DART spacecraft and mission designs have occurred since the previous descriptions of the kinetic impactor experiment (Cheng et al. 2016) and science investigations (Cheng et al. 2018). Most important is the addition of an Italian cubesat, called LICIACube, to the DART mission in 2018. LICIACube is contributed to DART by the Italian Space Agency (ASI). In addition, the mission design has changed in two aspects from that of Cheng et al. (2018): 1. DART is no longer a commercial rideshare but will have a dedicated launch on a Falcon 9 launch vehicle; and 2. The DART trajectory is a direct transfer to Didymos without a flyby of another object.

This paper will highlight updates to earlier descriptions of the DART mission (Cheng et al. 2016 and 2018) and will discuss two new aspects of the science investigations not covered in the earlier papers: 1. determination of momentum transfer efficiency $\beta$, and 2. contributions of LICIACube observations to estimation of $\beta$. LICIACube will image the DART impact ejecta plume and study its evolution, and it will image the non-impact hemispheres of Didymos. This paper will develop a model of the ejecta plume opacity as viewed from LICIACube. This model will not be applicable to impacts into volatile-rich targets, like Deep Impact at comet 9P/Tempel 1 and possibly also LCROSS at the Moon.

## 2. DART mission with LICIACube

The DART mission will demonstrate asteroid deflection by impacting the moon of the Didymos binary system in order to change the binary orbital period (Cheng et al. 2016, 2018). This change will be measured by Earth-based optical and radar observations. Optical light curve observations will measure the period change via the timing of mutual events (Scheirich and Pravec 2009), while radar observations will determine the primary shape and rotation as well as the orbital period and semi-major axis (Naidu et al. 2020). These measurements determine the orbital velocity change resulting from the DART impact. Since Didymos in September-October 2022 will approach within 0.072 AU from Earth, the optical light curve observations can use telescope apertures as small as 1 m.

Cheng et al. (2016) discussed the kinetic impactor experiment and predicted the changes in the binary orbit period, eccentricity and inclination with a simple 2-body Keplerian model of the binary system. Cheng et al. (2016) also presented an analytic model of the DART impact to predict the DART crater radius and total momentum transfer to the Didymos moon, using the point source impact scaling laws of Housen and Holsapple (2011). The analytical expression for the momentum transfer given by Cheng et al. (2016, eq. 9) was tested by Raducan et al. (2019), who compared the analytic predictions to the momentum transfer from numerical simulations, finding agreement within ~10%.

Both the DART spacecraft and mission designs have changed from those of Cheng et al. (2016). The DART incident momentum has changed, and the impact outcomes need to be updated accordingly. Also Cheng et al. (2016, Fig. 2 caption) stated incorrectly that the DART impact would be targeted so as to increase the Didymos orbital period. DART plans to impact the Didymos moon at elongation from the primary while the moon moves toward DART, so the impact will decrease the orbital period.

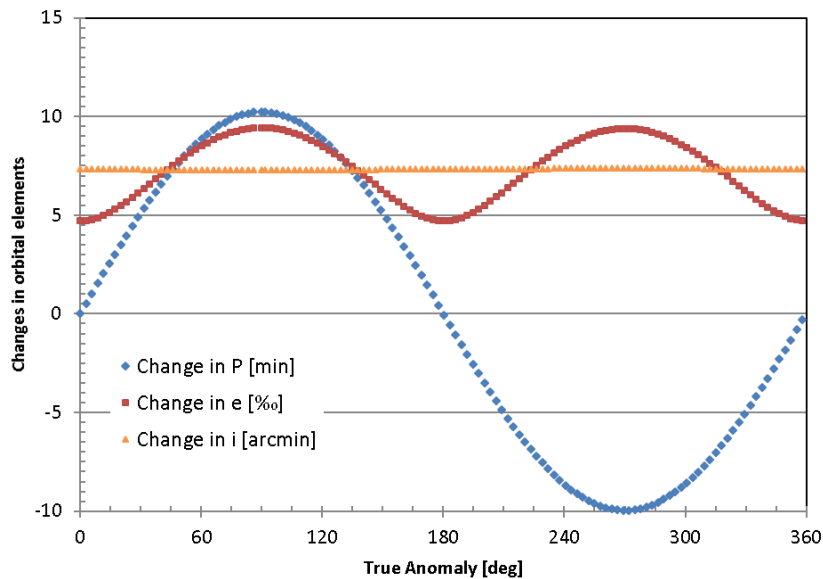

Figure 1. Changes in Didymos binary orbit period, eccentricity, and inclination after DART impact. DART plans to target true anomaly of 270° and reduce the orbit period.

The updated predictions of changes in the Didymos orbit after the DART impact are given in Fig. 1, assuming unit momentum transfer efficiency $\beta = 1$ and Keplerian motion, with a Didymos system mass $5.28\times10^{11}$ kg and a secondary mass of $4.8\times10^{9}$ kg as in Cheng et al. (2016). The binary orbit before the DART impact is assumed to be circular, consistent with Didymos light curve observations (Scheirich and Pravec 2009). Only the DART impulse component along the orbital velocity causes an orbital period change. DART will target true anomaly of 270°, where the impact will decrease the orbit period by ~10 min and will also increase the eccentricity to 0.009. The estimated DART spacecraft mass at impact is 610 kg.

A full 2 body dynamical model, which accounts for shapes and sizes of both bodies, is needed for quantitative description of the fully coupled rotational and translational motions of the system. These models show that, in addition to changes in period, eccentricity and inclination similar to those in Fig. 1, the DART impact will excite librations of Didymos B with amplitude up to ~10° (Cheng et al. 2018, Michel et al. 2018). The Hera mission plans to make detailed measurements of the librations and eccentricity after the DART impact to characterize this excited dynamical state (Michel et al. 2018). The excitation of librations will necessarily occur even if the DART impact is directed precisely through the target center of mass, because the impact changes the orbital period and the eccentricity. A DART impact slightly off center will induce both free and forced librations of Didymos B. Hera observations of these librations may determine moments of inertia of Didymos B and thereby constrain the internal structure (Michel et al. 2018).

Cheng et al. (2018) presented DART mission objectives, measurements and science investigations. The DART investigations fall into five working groups: impact simulations, ground-based observations of Didymos, dynamical modeling of Didymos, DART terminal approach imaging, and DART impact ejecta modeling. These are still the DART investigations after adding the Italian cubesat LICIACube, although the scope of the investigations is increased by inclusion of LICIACube observations. The updated DART mission design is summarized in Table 1, which includes a summary of the LICIACube flyby trajectory.

*Table 1. DART mission design with LICIACube flyby*

|  |  |  |
|---|---|---|
| DART | Launch Date | 22 July 2021 |
|  | Arrival Date | 30 September 2022 |
|  | Arrival Relative Speed | 6.58 km/s |
|  | Maximum Earth Distance | 0.1434 AU |
|  | Earth Distance at Impact | 0.072 AU |
|  | Solar Distance | 0.939 AU – 1.04 AU |
|  | Arrival Solar Phase Angle | 59.6° |
|  | Impact Angle to Orbit Plane | 15.5° |
| LICIACube | Release from DART | 5 days before impact |
|  | Flyby Speed Relative to Didymos | 6.58 km/s |
|  | Closest Approach Distance to Didymos | 55.4 km |
|  | Closest Approach Delay from DART Impact | 165.4 s |
|  | Time for which PL1 images are <5m/px | 58.4 s |

LICIACube is a 6U+ cubesat based on the ArgoMoon cubesat developed by Argotec and manifested to fly to the Moon on the NASA EM-1 mission. Like ArgoMoon, LICIACube is 3-axis stabilized and solar powered. After release 5 days prior to the DART impact, LICIA Cube will perform a small separation maneuver using a cold gas propulsion system, autonomously acquire Didymos with its imagers, and then perform a flyby of Didymos with closest approach 165.4 s after the DART impact. The LICIACube flyby speed is approximately the same, to within 5 m/s, as the DART impact speed on Didymos, and the closest approach distance to Didymos is 55.4 km. On this flyby trajectory, LICIACube will track Didymos throughout the approach and departure, including closest approach. The time delay of closest approach after the DART impact enables LICIACube to acquire images of the ejecta plume to study its structure and evolution. After closest approach, LICIACube images the non-impact hemispheres of Didymos A and B. Then for a period of several months after the flyby, LICIACube will return data directly to Earth via an X-band link.

The LICIACube imaging payload consists of two cameras. First is the PL1 camera, which is a monochrome imaging system with a 7.56 cm aperture, f/ 3 telescope. The PL1 field-of-view (FOV) is 2.9°×2.9°, and the instantaneous field-of-view (IFOV) is 5 arcsec/pixel. The images will study ejecta plume evolution, including the slower velocity ejecta fraction (at <5 m/s) which can be important for momentum transfer. Second is the PL2 camera, which is an RGB color imager with a wider FOV of 9.2°×4.9° and IFOV of 16 arcsec/px. This paper will focus on the PL1 measurements.

*Table 2. DART Objectives and Measurements*

| Objectives | Measurements |
|---|---|
| Demonstrate asteroid deflection by kinetic impact | Earth-based observations to determine orbital period change of the Didymos binary system induced by DART impact |
| Impact moon of Didymos system in Sept.-Oct. 2022 | |
| Determine the amount of deflection | |
| Determine momentum transfer efficiency of kinetic impact on an asteroid | DART approach imaging for impact site location and local surface geology |
| | DART approach imaging to measure sizes and shapes of Didymos A and B |
| Improve modelling and simulation capabilities | LICIACube imaging of impact ejecta plume density structure and evolution |
| | LICIACube imaging of Didymos A and B non-impact hemispheres |

Table 2 summarizes DART mission objectives and measurements. The primary objectives of the DART mission are to:

- demonstrate asteroid deflection by a spacecraft kinetic impact on the secondary member of the binary asteroid 65803 Didymos during its September-October, 2022 close approach to Earth,
- measure the amount of deflection by obtaining ground-based observations to determine the binary orbit period change,
- obtain a measure of the momentum transfer efficiency $\beta$ accounting for surface characteristics of the impact site, including physical properties like strength and porosity as well as surface slopes and structural features.

The ground-based telescopic observations are the primary DART observations to measure the asteroid deflection from the kinetic impact by determining the change in orbital speed. Combined with measurements of the Didymos B size and shape from DART approach imaging, the mass of Didymos B is estimated and then the momentum transfer. The determination of momentum transfer efficiency $\beta$ is then improved with the aid of numerical simulations, using DART approach imaging observations and LICIACube flyby imaging. This approach to estimation of $\beta$ from DART datasets and simulations will be further discussed below.

DART approach imaging will determine the DART impact location and characterize the local surface slope and structural features like blocks. Both surface slopes (Bruck Syal et al. 2016; Feldhacker et al. 2017) and blocks that may be present at or near the impact site (Owen et al. 2017, Stickle et al. 2018) can affect the momentum transfer efficiency $\beta$. Large blocks near the artificial impact site on [162173] Ryugu influenced the cratering process there (Arakawa et al. 2020). The DART approach images are acquired by the DRACO camera which uses a 20.8 cm aperture, f/ 12.6 telescope with FOV of 0.29°×0.29°. The DRACO IFOV is 0.5 arcsec/px, and terminal approach images just before impact will achieve a ground sampling distance of 50 cm per pixel or better, enabling characterization of surface features down to the size of the DART spacecraft.

This paper will discuss how DART will obtain a measure of $\beta$, the primary planetary defense objective (Holsapple and Housen 2012). The approach to estimation of $\beta$ has not been discussed previously. This paper will present a new model of the ejecta plume evolution as viewed from LICIACube, and it will discuss how these plume images can discriminate the differences in plume structure and evolution that result from different target physical properties (mainly strength and porosity), thereby allowing constraints on these properties which significantly affect $\beta$ (Stickle et al. 2015, Bruck Syal et al. 2016).

The next section will discuss momentum transfer efficiency $\beta$ and its estimation from DART datasets and impact simulations. The following section will present a model of the ejecta plume imaged by LICIACube and discuss inference of target properties.

## 3. Momentum Transfer Efficiency

When a kinetic impactor of mass $m$ strikes a target at a velocity $\boldsymbol{U}$, the momentum transferred to the target of mass $M$, written as $M\Delta\boldsymbol{v}$, can exceed the incident momentum $m\boldsymbol{U}$ because of momentum carried away in a backward direction by impact ejecta. The momentum transfer efficiency $\beta$ is defined by (Feldhacker et al., 2017)

$$M\Delta\boldsymbol{v} = m\,\boldsymbol{U} + m(\beta - 1)(\boldsymbol{n} \cdot \boldsymbol{U})\boldsymbol{n} \qquad [1]$$

using the outward surface normal unit vector $\boldsymbol{n}$ at the impact site. In eq. 1, the first term on the right is the incident momentum and the second term is the momentum contribution of escaping impact ejecta, which is assumed to be along the surface normal vector.

This definition of $\beta$ can be re-expressed as the ratio of the normal components of the momentum transfer and the incident momentum, or

$$\beta = \frac{M(\boldsymbol{n}\cdot\Delta\boldsymbol{v})}{m(\boldsymbol{n}\cdot\boldsymbol{U})} \qquad [2]$$

In general, the vector momentum transfer $\Delta M\boldsymbol{v}$ is not collinear with the incident momentum vector $m\boldsymbol{U}$ because of the ejecta momentum vector, which is not anti-parallel to the incident direction, but is affected by either or both of: 1. the local surface inclination to the incident direction as discussed by (Feldhacker et al. 2017), or 2. presence of topography or a blocky surface at the impact site (Owen et al. 2017, Stickle et al. 2018). For the DART impact, approach imaging will provide information on these impact conditions and target characteristics (Cheng et al. 2018), by determining the impact location and the local surface inclination, and identifying few meter scale or larger blocks if present. LICIACube ejecta plume imaging will constrain or determine the direction of the ejecta momentum. If this direction is affected by local topography and/or blocks, it will be compared to the surface normal vector defined as an average over the pre-impact surface.

The primary measurements of asteroid deflection made by the DART mission are the ground-based telescopic measurements of the orbital period change $\Delta P$ from the DART impact. The period change measurement determines only the transverse velocity change $\Delta v_t$ (the component along the circular orbit motion). The other two components of $\Delta \boldsymbol{v}$ are not measured by DART. The period change will, in the approximation of circular Keplerian motion, determine the efficiency of impact kinetic energy transfer to circular orbit energy $E$ by $\frac{\Delta P}{P} = -\frac{3\Delta E}{2E}$. However, only the transverse component of the momentum transfer $M\Delta v_t$ is determined from the transverse velocity change, given the target mass $M$.

The DART impact geometry from Table 1 and Fig. 1 is such that the incident momentum $m\boldsymbol{U}$ is directed nearly opposite, at an angle of 164.5°, to the direction of orbital motion $\boldsymbol{e}_t$. If it is assumed that only the component of incident momentum along the orbital direction contributes to the transverse momentum transfer, then $\beta$ is estimated by

$$\beta \cong \frac{M \Delta v_t}{m\boldsymbol{U} \cdot \boldsymbol{e}_t} \qquad [3]$$

where the orbital period change determines $\Delta v$ by $\frac{\Delta v}{v} = -\frac{\Delta P}{3P}$ in the Keplerian approximation. DART will use full 2-body dynamical models to make an accurate determination of $\Delta v$ (Cheng et al. 2018). DART will determine the target body mass $M$ from approach imaging of the size and shape and hence the volume, assuming that the Didymos primary bulk density 2100 kg m$^{-3}$ (Michel et al. 2016) applies also to the secondary. In addition, LICIACube will provide images of the non-impact hemisphere of Didymos B obtained after closest approach, viewing the side of Didymos not seen by DART. LICIACube images will significantly improve the volume determination for Didymos B. An initial estimate for $\beta$ is then obtained from eq. [3].

LICIACube will obtain additional observations of the ejecta plume structure and evolution in order to constrain and infer the ejecta momentum $\boldsymbol{p}_{ej}$ both in direction and magnitude, providing important information to help determine $\beta$. A model of the ejecta plume evolution as imaged by LICIACube will be presented in the following sections and applied to a discussion of how observed differences in plume structure and evolution can discriminate between different target physical properties (mainly strength and porosity), thereby allowing inferences on $\boldsymbol{p}_{ej}$. Plume images further provide direct information on the direction of $\boldsymbol{p}_{ej}$.

DART will use numerical simulations of the kinetic impact (Cheng et al. 2018, Stickle et al. 2020) to help determine $\beta$ and to understand uncertainties arising from dependencies on impact conditions and target characteristics. DART will develop numerical simulations which match DART input data and observations of impact outcomes, including the following: 1. Spacecraft of final mass $m$ with known spacecraft structure impacting at velocity vector $\boldsymbol{U}$; 2. Transverse component of momentum transfer consistent with the observed period change; 3. Impact on target body shape at observed location matching local surface inclination and topographic and structural features such as boulders down to 2 m scale; 4. Strength, damage, failure models and porosity models consistent with imaging from DART and LICIACube of ejecta plume and target body. These numerical simulations of the impact, which predict the observed $M\Delta\boldsymbol{v}$ transverse component, can then be used to calculate the other two components of $M\Delta\boldsymbol{v}$, and then $\beta$ can be determined from eq. [1]. Pathways from DART datasets to determinations or estimates of $\beta$ are illustrated in Fig. 2, showing the important role of numerical simulations.

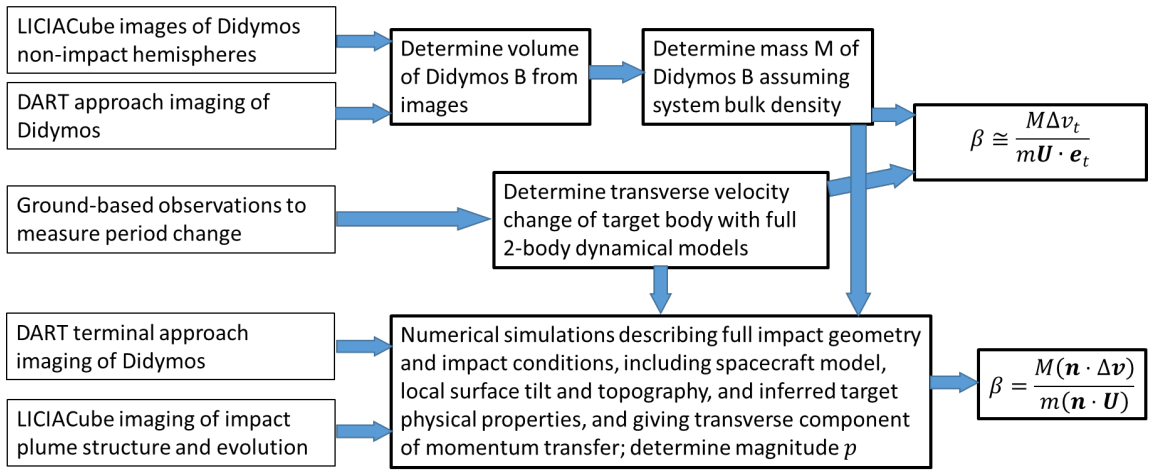

Figure 2. Pathways from DART datasets (left side) to determinations of $\beta$ (right side).

## 4. DART Impact Ejecta Plume

This section will present a model for the DART impact ejecta plume as observed by LICIACube from its fast flyby of Didymos (Table 1) using the PL1 camera. The flyby images will be acquired from a fast moving platform, such that initial images are obtained looking down roughly along the plume axis at the impact site from great distance in back scatter geometry; then images are acquired at close range looking nearly perpendicular to the plume axis after a sufficient time delay to observe even slow moving ejecta at altitude; then images are acquired after closest approach looking again roughly along the plume axis in forward scatter geometry (here the impact site is behind Didymos B and the plume above the limb can be imaged). The LICIACube images of the plume density structure and evolution provide information on the ejecta mass versus velocity distribution, because that distribution determines how much mass reaches a given altitude at a given time. The ejecta mass versus velocity distribution depends on target physical properties, so that constraining the ejecta distribution also constrains these properties.

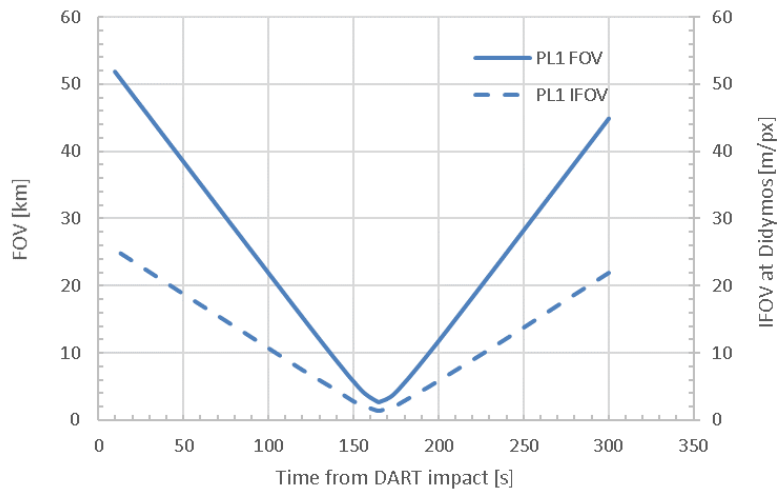

Figure 3 The PL1 imager FOV (left axis) and IFOV (right axis) at Didymos. PL1 FOV is the full width of the FOV. PL1 IFOV is the resolution as given by ground sampling distance.

The width of the PL1 FOV at the distance of Didymos is shown in Fig. 3, for imaging from the LICIACube flyby trajectory, versus time after the DART impact. LICIACube closest approach occurs at 165.4 s after the DART impact, and the FOV is 2.8 km wide at that time. Ejecta released at a typical speed of 5 m/s (Jutzi and Michel 2014) will have reached an altitude ~ 800m at that time, and even faster ejecta remain within the FOV. Fig. 3 also shows the PL1 ground sampling distance at Didymos versus time. The sampling distance is 5 m/px or better for a time interval of 58.4 s, and at closest approach it is 1.4 m/px.

The ejecta plume model is based upon point source impact scaling laws of Housen and Holsapple (2011) as applied by Cheng et al. (2016) to the DART impact. A spherical impactor of mass $m$ and radius $a$ is incident normally at velocity $U$. The crater radius $R$ is non-dimensionalized by target density $\rho$ and impactor mass $m$, and the combination $R\left(\frac{\rho}{m}\right)^{1/3}$ is expressed in terms of the dimensionless scaling parameters $\pi_2 = \frac{ga}{U^2}$ and $\pi_3 = \frac{Y}{\rho U^2}$ in the gravity- or strength-controlled impact cases, respectively. Here $\pi_2$ is the gravity-scaled size, with $g$ the target surface gravity; and $\pi_3$ is the strength parameter (the ratio of material strength and inertial stresses) with target impact strength $Y$ and target density $\rho$. In the gravity-controlled case, the crater radius $R$ is given by

$$R\left(\frac{\rho}{m}\right)^{1/3} = H_1 \left(\frac{\rho}{\delta}\right)^{\frac{2+\mu-6\nu}{3(2+\mu)}} \left(\frac{ga}{U^2}\right)^{-\frac{\mu}{2+\mu}} \quad \text{(gravity)} \quad [4a]$$

In the strength-controlled case the crater radius $R$ is given by

$$R\left(\frac{\rho}{m}\right)^{1/3} = H_2 \left(\frac{\rho}{\delta}\right)^{(1-3\nu)/3} \left(\frac{Y}{\rho U^2}\right)^{-\mu/2} \quad \text{(strength)} \quad [4b]$$

The dimensionless scaling parameter $\mu$ depends on target properties and lies in the range $1/3 < \mu < 2/3$, where $\mu = 1/3$ is the momentum scaling limit, and $\mu = 2/3$ is the energy scaling limit. The scaling parameter $\nu$ enters via the ratio of target to projectile densities $\rho/\delta$, and $\nu$ is empirically about 0.4 for any target material. The normalization of crater size (and thus total ejecta mass) is given by $H_1$ or $H_2$ for gravity or strength scaling, respectively.

Empirical values for these parameters, based on fitting to ejecta distributions from laboratory experiments (Housen and Holsapple, 2011), are shown in Table 3 for four target cases with strength-controlled impacts, labeled C2, C3, C7 and C8, and one gravity-controlled case labeled C5. Numerical simulations (Prieur al. 2017) have determined appropriate scaling parameters for additional target materials spanning a wide range of target properties.

*Table 3. Target cases for DART impact ejecta modeling*

| Target* | Porosity | $\mu$ | $C_1$ | $k$ | $H_1, H_2$ | $p$ | $a$ (mm) | $U$ (m/s) | $\delta$ (kg/m3) | $\rho$ (kg/m3) | $Y$ (MPa) |
|---|---|---|---|---|---|---|---|---|---|---|---|
| Basalt C2 | ~0 | 0.55 | 1.5 | 0.3 | 1.1 | 0.5 | 1.6 | 6200 | 2700 | 3000 | 30 |
| WCB C3 | 20% | 0.46 | 0.18 | 0.3 | 0.38 | 0.3 | 3.6 | 1860 | 2700 | 2600 | 0.45 |
| SFA C7 | 45% | 0.4 | 0.55 | 0.3 | 0.4 | 0.3 | 7 | 1900 | 930 | 1500 | 0.004 |
| PS C8 | 60% | 0.35 | 0.6 | 0.32 | 0.81 | 0.2 | 8.7 | 1800 | 940 | 1200 | 0.002 |
| Sand C5 | 35% | 0.41 | 0.55 | 0.3 | 0.59 | 0.3 | 3.9 | 6770 | 1220 | 1510 | ** |

*WCB = weakly cemented basalt; SFA = sand/fly ash; PS = perlite/sand; C2, C3, C5, C7 and C8 are labels used by Housen and Holsapple (2011); in strength-controlled cases, $\nu = 0.4, n_1 = 1.2, n_2 = 1$.

**gravity-controlled case C5, where $\nu = 0.4, n_1 = 1.2, n_2 = 1.3$.

The target cases in Table 3 are arranged in descending order of target strength, from the strong intact basalt case that is unlikely to apply (Cheng et al. 2016), to strength-controlled cases with strengths $Y$ as low as a few kPa that may be most relevant to Didymos, and to finally a gravity-controlled case with negligibly small strength. For the DART impact, assuming the Didymos density of 2.1 g cm$^{-3}$ for the target body (Michel et al. 2016) and assuming the projectile radius $a = 686$ mm, the gravity scaling case applies if the strength $Y < 4.7$ Pa. It is likely that the target strength is high enough that strength scaling applies; even a strength as low as for lunar regolith $Y \approx 1$ kPa would be easily strength-controlled at the scale of the DART impact which is typical for asteroid mitigation by kinetic impact (Cheng et al. 2016; Holsapple and Housen 2012). The upper surface strength on Comet 67P inferred from the Philae landing (Biele et al. 2015) was estimated as ~1 kPa, although geologic features on Comet 67P (overhangs) required tensile/shear strengths of only 10 Pa to 20 Pa (Thomas et al. 2015). Cometary meteoroid strengths (from bolide break-up in the upper atmosphere) are also typically ~ 1 kPa (Trigo-Rodriguez and Llorca 2006). For the DART impact to be gravity controlled, the target strength must be lower than any of these, but a gravity-controlled case is included in addition to the strength-controlled cases considered by Cheng et al. (2016). The Hayabusa 2 artificial impact on Ryugu produced a gravity-controlled crater (Arakawa et al. 2020).

The ejecta plume mass versus velocity distribution is described following Housen and Holsapple (2011). The speed of ejecta $v$ that are released at radial distance $x$ from the central point of impact, when non-dimensionalized by the incident velocity $U$, is

$$\frac{v}{U} = C_1 \left[\frac{x}{a}\left(\frac{\rho}{\delta}\right)^\nu\right]^{-1/\mu} \left(1 - \frac{x}{n_2 R}\right)^p \qquad [5\,a]$$

The impact is at normal incidence, and ejecta are released on ballistic trajectories. The mass $M$ ejected from within $x$, which is also the mass ejected above the corresponding speed according to eq. (5a), is

$$\frac{M}{m} = \frac{3k}{4\pi}\frac{\rho}{\delta}\left[\left(\frac{x}{a}\right)^3 - n_1^3\right] \qquad [5\text{ b}]$$

The ejecta mass versus velocity distribution is defined implicitly by eqs. (5a) and (5b) and has a cutoff at high ejecta velocity, corresponding to the cutoff at small $x = n_1 a$ in eq. (5b), and additionally a cutoff at low ejecta velocity corresponding to large $x = n_2 R$ from eq. (5a). The parameters $C_1$ and $k$ normalize ejecta velocities and ejecta mass, respectively.

An alternative scaling law for the ejecta velocity $v(x)$ has been proposed by Raducan et al. (2019), in which eq. (5a) is replaced by a form with cutoffs appearing at both small $x = n_1 a$ and large $x = n_2 R$. The scaling law eq. (5a) is used here, because according to experiment the ejecta velocity maximizes at small $x$.

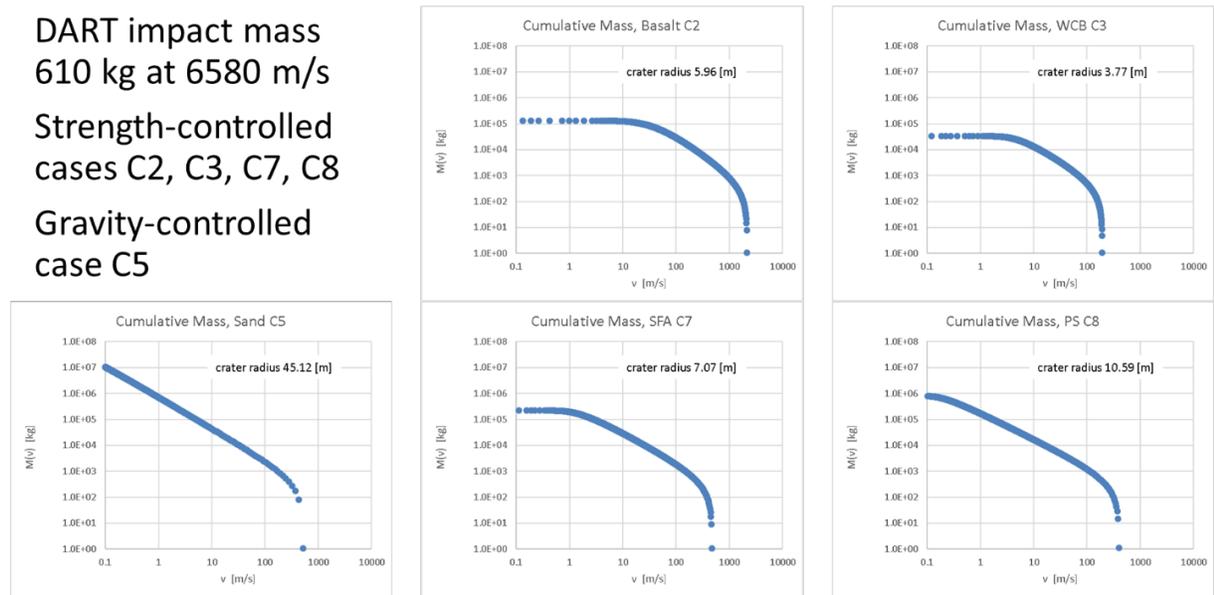

Figure 4 DART impact ejecta cumulative mass distributions versus velocity, for five target cases.

Fig. 4 shows the cumulative distributions of mass $M(v)$ ejected above velocity $v$ as found from eqs. (5a, 5b) for the five target cases in Table 3. The high strength case basalt C2 has the most high speed ejecta. The very weak target cases, C7, C8 and C5 in order of decreasing strength, have the most low speed ejecta, and the lower the strength the greater the proportion of the slowest ejecta. Since slower ejecta need longer time to rise to a given altitude, these differences in mass versus velocity distribution directly affect the ejecta plume structure and evolution as imaged by LICIACube at a particular time.

The total momentum carried to infinity by impact ejecta is given by an integration over $w = x/a$ (Cheng et al. 2016),

$$p_{ej} = \frac{9km}{4\pi} \frac{\rho}{\delta} \int_{n_1}^{n_2 R/a} dw \, w^2 \, v_{inf} \cos\theta \quad [6]$$

where $v_{inf} \cos\theta$ is the asymptotic velocity component along the incidence direction, and the momentum transfer efficiency is found from eq. [2] by numerical quadrature. Cheng et al. (2016) gave an analytic approximation for $\beta$ in strength-controlled impacts,

$$\beta - 1 \cong \frac{9kC_1}{4\pi\sqrt{2}} \left(\frac{\rho}{\delta}\right)^{(\mu-\nu)/\mu} \frac{\mu}{3\mu-1} \{(0.74 n_2 \, R/a)^{(3\mu-1)/\mu} - n_1^{(3\mu-1)/\mu}\} \quad [7]$$

which agrees to ~10% with results of iSALE numerical simulations for impact conditions similar to those of the DART impact (Raducan et al. 2019).

*Table 4 DART Impact Outcomes from Scaling Laws*

|  | Basalt C2 | WCB C3 | SFA C7 | PS C8 | Sand C5 |
|---|---|---|---|---|---|
| Momentum transfer efficiency $\beta$ by eq. [2] | 3.05 | 1.09 | 1.28 | 1.22 | 1.89 |
| Crater radius $R$ [m] | 5.96 | 3.77 | 7.07 | 10.6 | 45.1 |

Table 4 shows the predicted $\beta$ from numerical integrations of eq. [6] and the crater radius $R$ for five target cases (Housen and Holsapple 2011), assuming a DART impact mass of 610 kg incident at 6580 m/s. The highest $\beta$ is predicted for the basalt case C2, and the lowest $\beta$ is for the weakly cemented basalt case C3; the gravity-controlled case sand C5 gives the second highest $\beta$. However, the gravity-controlled case predicts a crater diameter not much less than the target body radius, in which case target body curvature may be important. Since target curvature is not accounted for in the scaling laws, the predicted crater radius $R$ is uncertain in this case, which is included to show the behavior of the ejecta plume structure and evolution with a gravity controlled impact.

4.1. Impact Ejecta Plume as Observed by LICIACube

This section develops a model for the ejecta plume opacity as imaged by the PL1 camera during the LICIACube flyby of Didymos. LICIACube will point PL1 to Didymos autonomously throughout the approach, closest approach, and early departure phases of the flyby. Plume images can be acquired at any of these times.

The ejecta are assumed to be released from the impact point at an angle $\alpha = 45°$ to the target surface (see Fig. 4), filling a hollow cone with ejection angles in the small range $(\alpha, \alpha + \delta\alpha)$. That is, trajectories are calculated assuming that all ejecta are released not only at a fixed angle $\alpha$ but also from a single point, neglecting the variation in release points within the crater radius $R$ compared to the range to LICIACube (1088 km at DART impact down to 55.4 km at closest approach), for the purpose of locating the ejecta as a function of time after the DART impact in PL1 images. The variation in ejection angles over roughly 40°-60° (Hermalyn et al. 2012, Luther et al. 2018, Gulde et al. 2018) is also neglected. A further simplification is made that only ejecta well above the escape speed (~10 cm/s) are considered, since these ejecta make the dominant contribution to $\beta$ (Jutzi and Michel 2014; Cheng et al. 2016). The model represents these ejecta as being released on rectilinear trajectories at constant speed, an excellent approximation for ejecta that have not moved much more than a km from the impact site. However, the time of ejecta release is accounted for as in point source scaling (O'Keefe and Ahrens, 1993), according to $t_{release} = \frac{a}{U}\left(\frac{x}{a}\right)^{(1+\mu)/\mu}$.

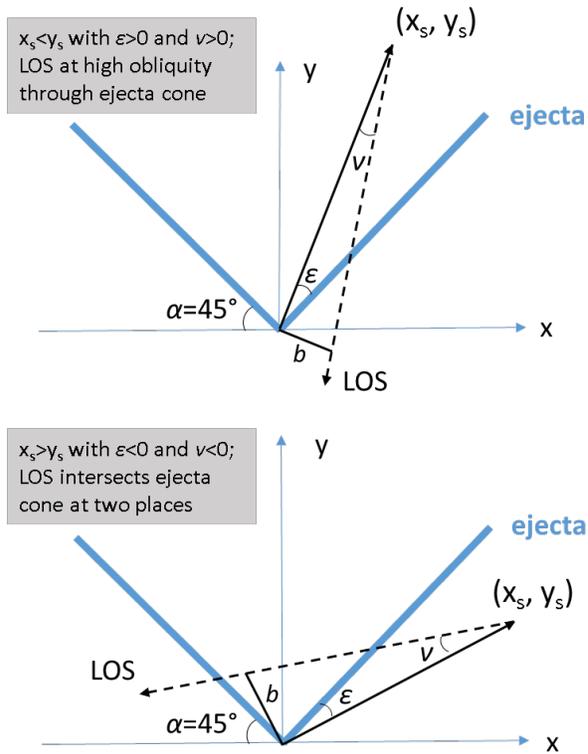

Figure 5 Viewing geometries from LICIACube along flyby trajectory: during early approach (upper panel) LICIACube is within the ejecta cone $x_s < y_s$ and LOS can intersect the ejecta cone at high obliquity as shown or at low obliquity; after crossing outside ejecta cone (lower panel) where $x_s > y_s$ a single LOS intersects the ejecta cone twice.

Since LICIACube follows DART on almost the same trajectory as DART, LICIACube on approach viewing Didymos is also viewing the impact site almost at normal incidence, assuming DART hits its target in the center. The plume opacity profiles will be calculated in an x-y plane (see Fig. 5) defined as the plane containing the axis of the 45° ejecta cone and the vector to LICIACube from the impact site. The origin is at the impact site and the y-axis is the axis of the ejecta cone. LICIACube is at position $(x_s, y_s)$. The flyby trajectory is such that $x_s = 55.4$ km and $y_s$ decreases with time.

Fig. 5 illustrates the LICIACube viewing geometry at two times, early in the approach to Didymos where $x_s < y_s$ (inside ejecta cone, or $\varepsilon > 0$) and later after $y_s$ has decreased sufficiently to make $x_s > y_s$, meaning that LICIACube has crossed to the outside of the ejecta cone, $\varepsilon < 0$. LOS indicates the instantaneous line-of-sight for a pixel in a PL1 image, and opacity will be calculated as a function of the angle $\nu$ which is re-expressed as a projected distance in the plane-of-sky through the impact site and written as $b$. During the early approach, LOS intersects the ejecta cone at high obliquity $\gamma = 1/\sin(\varepsilon + \nu)$ for positive $\nu$ and at low obliquity $\gamma = 1/\cos(\varepsilon + \nu)$ for negative $\nu$. After LICIACube crosses the ejecta cone, a single LOS intersects the ejecta cone twice, and $\gamma = -1/\sin(\varepsilon + \nu)$ for the high obliquity intersection for negative $\varepsilon$ and $\nu$.

The ejecta plume optical depth is now calculated by an integration along the LOS. At the time of an image $t$ relative to the DART impact, the LOS for an image pixel intersects the ejecta cone once or twice (if twice, the two contributions are summed). If the LOS intersects the ejecta cone at radial distance $r$ from the impact site, the ejecta were released at velocity $v = r/t$. At the intersection point, the spherical volume element is written as $volume = 2\pi r^2 \sin\alpha \, \delta r \, \delta\alpha$ and the path length through the intersection is written as $path = \gamma \, r \, \delta\alpha$ where $\gamma > 1$ accounts for obliquity; ejecta angle $\alpha = 45°$ is shown in Fig. 4. The ejecta mass within the volume element is written as $mass \ in \ \delta r = \left|\frac{dM(v)}{dr}\right| \delta r$ which is found from a numerical differentiation of the cumulative mass distribution $M(v)$. The optical depth contribution from the intersection is then

$$optical \ depth \ \tau = \frac{\left|\frac{dM(v)}{dr}\right| \delta r \, \gamma \, r \, \delta\alpha}{2\pi r^2 \sin\alpha \, \delta r \, \delta\alpha} \left(\frac{area}{mass}\right) Q = \frac{\left|\frac{dM(v)}{dr}\right| \gamma}{2\pi r \sin\alpha} \left(\frac{area}{mass}\right) Q \qquad [8]$$

where the factor $\left(\frac{area}{mass}\right)$ is the total physical cross section per unit mass of ejecta; this quantity is given by the particle size distribution, to be discussed below. The factor $Q$ is the scattering or extinction efficiency, relating the physical cross section area to the scattering or extinction cross section, respectively.

The total cross sectional area per unit mass $\left(\frac{area}{mass}\right)$ is found from an assumed ejecta particle size distribution. An Itokawa size distribution is adopted, where the (differential) number of particles is $n(s) = 2.746 \times 10^5 \, s^{-3.98}$ in the size range from $d_1 = 0.001$ m to $d_2 = 1$ m. The area of particles is $A_d = \int_{d_1}^{d_2} ds \, \pi s^2 \, n(s)/4$ and the volume is $V_d = \int_{d_1}^{d_2} ds \, \pi s^3 \, n(s)/6$, from which $\left(\frac{area}{mass}\right) = \frac{A_d}{\rho V_d} = 0.1962$ m²/kg. This adopted size distribution is consistent with boulder size distributions on Itokawa (Tancredi et al. 2015; Mazrouei et al. 2014) at meter scale and larger, and also with the returned sample particle size distribution (Nakamura et al. 2012).

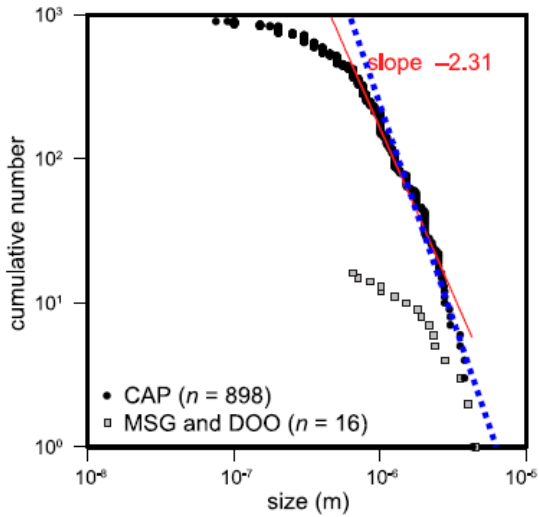

Figure 6. Itokawa returned sample size distributions (Nakamura et al. 2012), with solid line the cumulative number slope -2.31 reported by Nakamura et al., compared to a slope -3 (dotted blue line). CAP is a sample of solid fragments, MSG and DOO are samples of quenched melt droplets.

Fig. 6 reproduces figure 6 of Nakamura et al. with an additional line plotted at cumulative slope -3, consistent with the distribution adopted here, which compares well to the data for solid fragments larger than $10^{-6}$ m. The Itokawa size distribution may be consistent with re-accumulation from dispersed material, where fine particles under ~cm size were removed by solar radiation pressure unless they were stuck on the surfaces of much larger particles (Cheng et al. 2009). A cumulative size distribution slope ~ -3 is also consistent with impact experiments (Buhl et al. 2014).

4.2. LICIACube Plume Imaging Optical Depth Profiles

Optical depth profiles of $\tau$ versus the distance $b$ are calculated from eq. [8] and plotted in Figs. 7-15, where the pixel line-of-sight points at the impact site when $b = 0$. The profiles in Figs. 7-15 show optical depth from the physical cross section, meaning $Q=1$. The scattering and extinction efficiencies $Q$ depend not only on particle sizes but also on shape and composition (Fu and Sun, 2001). For an initial study of what can be obtained from plume imaging by LICIACube, particularly regarding estimation of momentum transfer efficiency, simplifications are proposed for $Q$. Namely, for estimating the obscuration of the target body surface as seen by PL1 through the plume, that is, for extinction optical depth in the visible, the approximation $Q_e \sim 2$ is adopted bearing in mind that the assumed smallest particle size $d_1 = 0.001$ m is much larger than the wavelength. For estimating the brightness of the plume seen against dark sky when the plume is optically thin, the rough estimate $I/F \sim 0.1\,\tau$ is adopted, where $I$ is scattered radiance and $\pi F$ is solar irradiance, and where $\tau$ is optical depth with $Q = 1$. This $I/F$ is equivalent to assuming a scattering phase function of 0.4 for plume particles large compared to an optical wavelength, observed at phase angles $\geq 60°$. The Didymos optical albedo is 0.15 (Naidu et al. 2020).

Figs. 7-15 compare the ejecta plume evolution imaged by LICIACube for five impact cases, one gravity-controlled and four strength-controlled, with differing target strengths and porosities. Fig. 7 shows the optical depth profiles obtained at $t$=1 sec, which is only one second after the DART impact at $t$=0, at a distance of 1082 km from Didymos, with PL1 image resolution of 26.76 m/px. Fig. 8 shows the profiles shortly afterward at $t$=10 sec, with LICIACube now closer to Didymos with PL1 resolution of 25.3 m/px. Comparison of Figs. 7 and 8 already brings out an important aspect of the plume evolution of the highest strength and lowest porosity case (basalt C2) compared to the other cases at lower strength and higher porosity. Namely, in case C2, clearing of the ejecta over the impact site (decrease in optical depth with time) can already be seen by 10 sec after the impact.

From Figs. 7 and 8, the plume optical depth for case C2 is seen to decrease between $t$=1 s and $t$=10 s within distance $|b| < 100$ m from the impact site, while the optical depth increases with time at larger $|b|$ as the ejecta move outward. For case WCB C3, clearing of the ejecta has actually started by $t$=10 s, but the decrease in optical depth occurs only within $|b| < 15$ m from the impact site, too small to be resolvable by PL1, and optical depth increases with time at larger $|b|$. For the lower strength and higher porosity cases C7, C8, and C5, there is no clearing over the impact site between $t$=1 s and $t$=10 s, because more of the ejecta are released at lower velocities.

Fig. 9 shows that for case WCB C3, the clearing of the ejecta over the impact site continues and is well resolvable in the PL1 images by $t$=20 s. Between $t$=10 s and $t$=20 s, clearing continues also for the basalt C2 case, but now out to $|b| < 400$ m; optical depth is rising with time at larger $|b|$. For the low strength and high porosity cases C7, C8, and C5, there is no clearing over the impact site and optical depth continues to increase at all $b$.

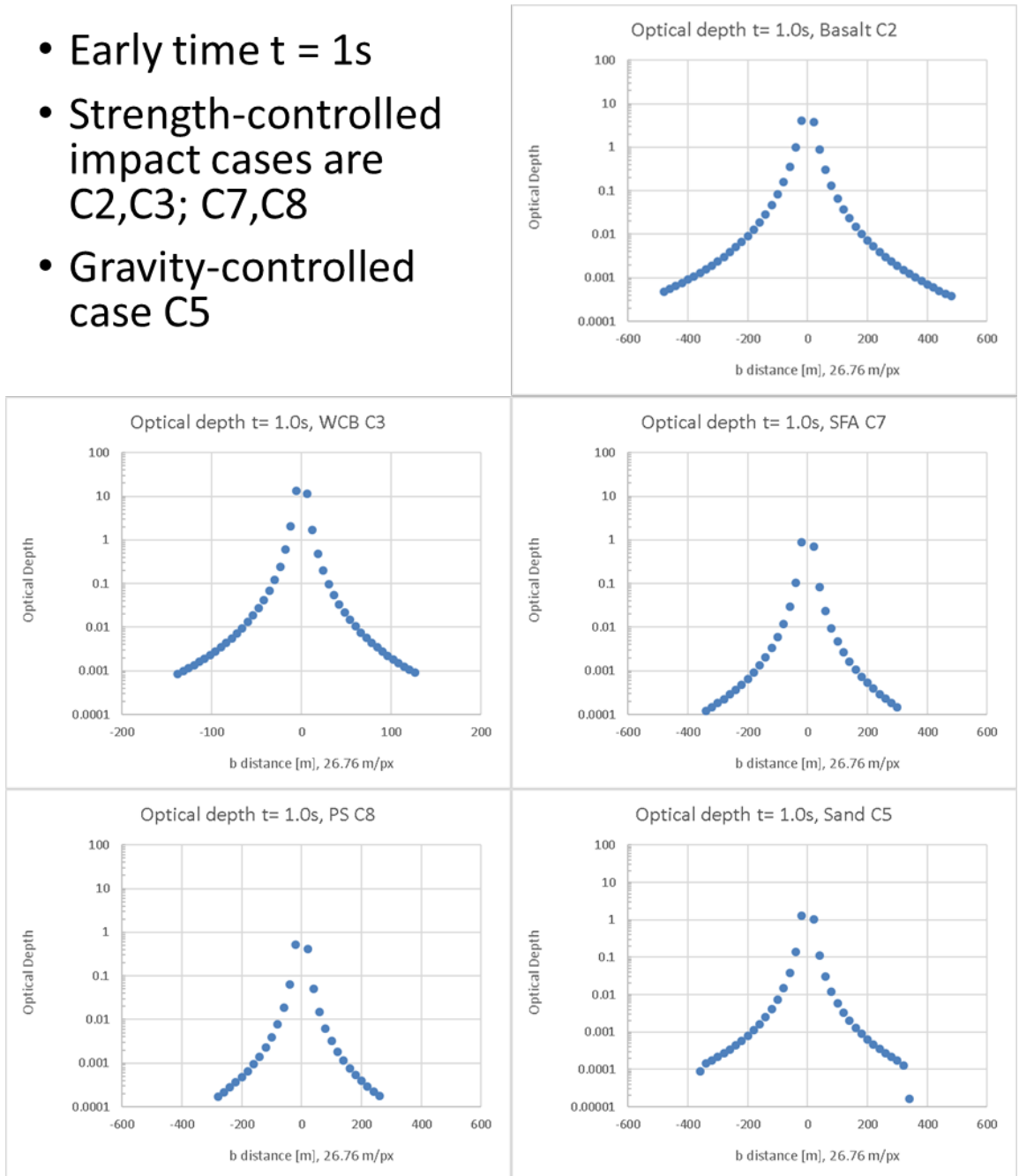

Figure 7. Optical depth profiles at time *t*=1 s after DART impact, plotted versus projected distance *b* at Didymos; DART crater at *b*=0, and target body radius ~81 m. Imager PL1 resolution is 26.76 m/px ground sampling distance at Didymos, and $\alpha + \varepsilon = 87.068°$; see Fig. 4

- Early time t = 10s
- Strength-controlled impact cases are C2,C3; C7,C8
- Gravity-controlled case C5

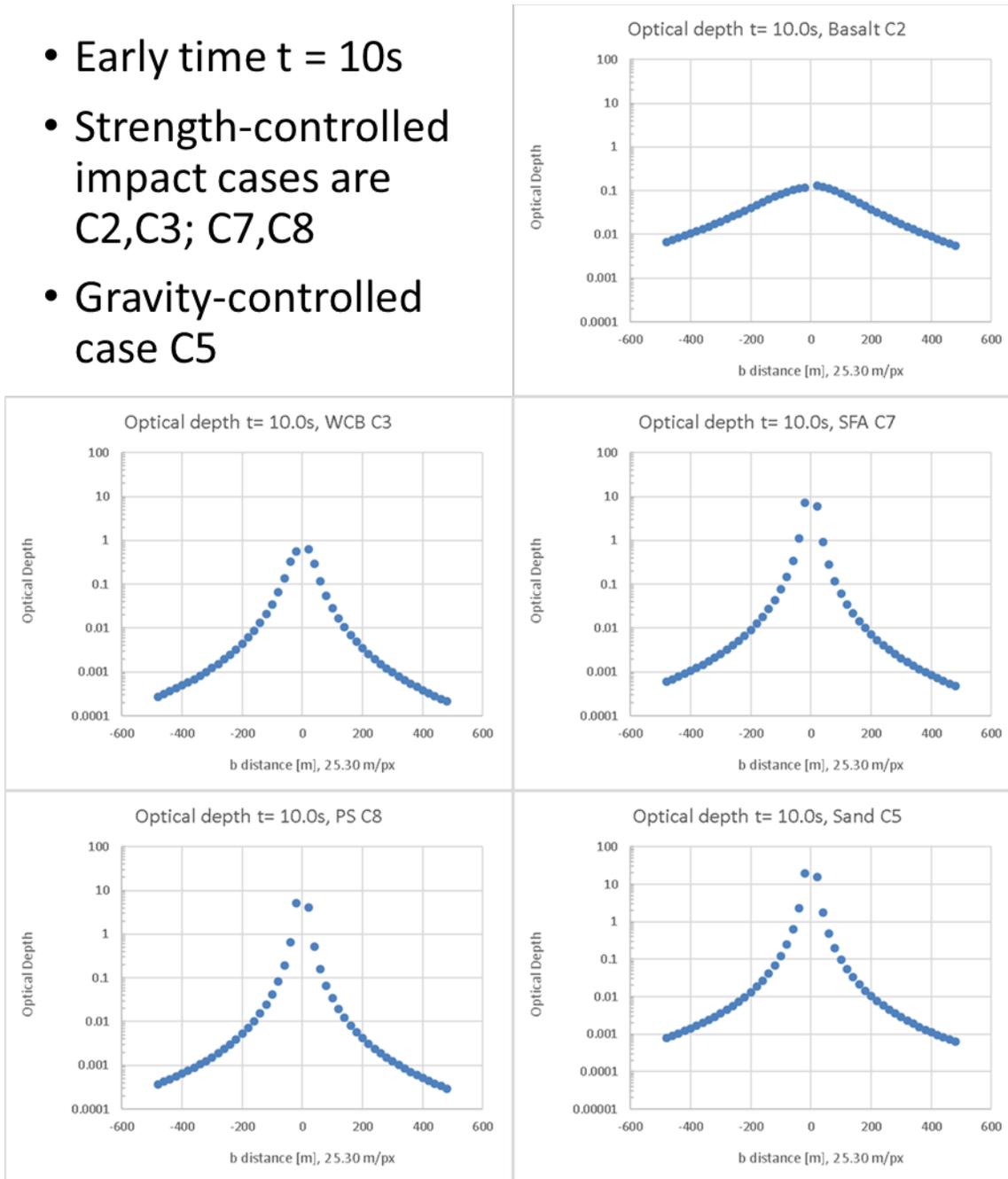

Figure 8. Same as Fig. 7, at time t=10 s after DART impact, where $\alpha + \varepsilon = 86.899°$. Compared to Fig. 7, clearing of ejecta near impact site for case C2, or decreasing optical depth $\tau$ for $|b| < 100$ m with increasing $\tau$ otherwise. In weak target cases (bottom row), $\tau$ is increasing with time at all distances plotted.

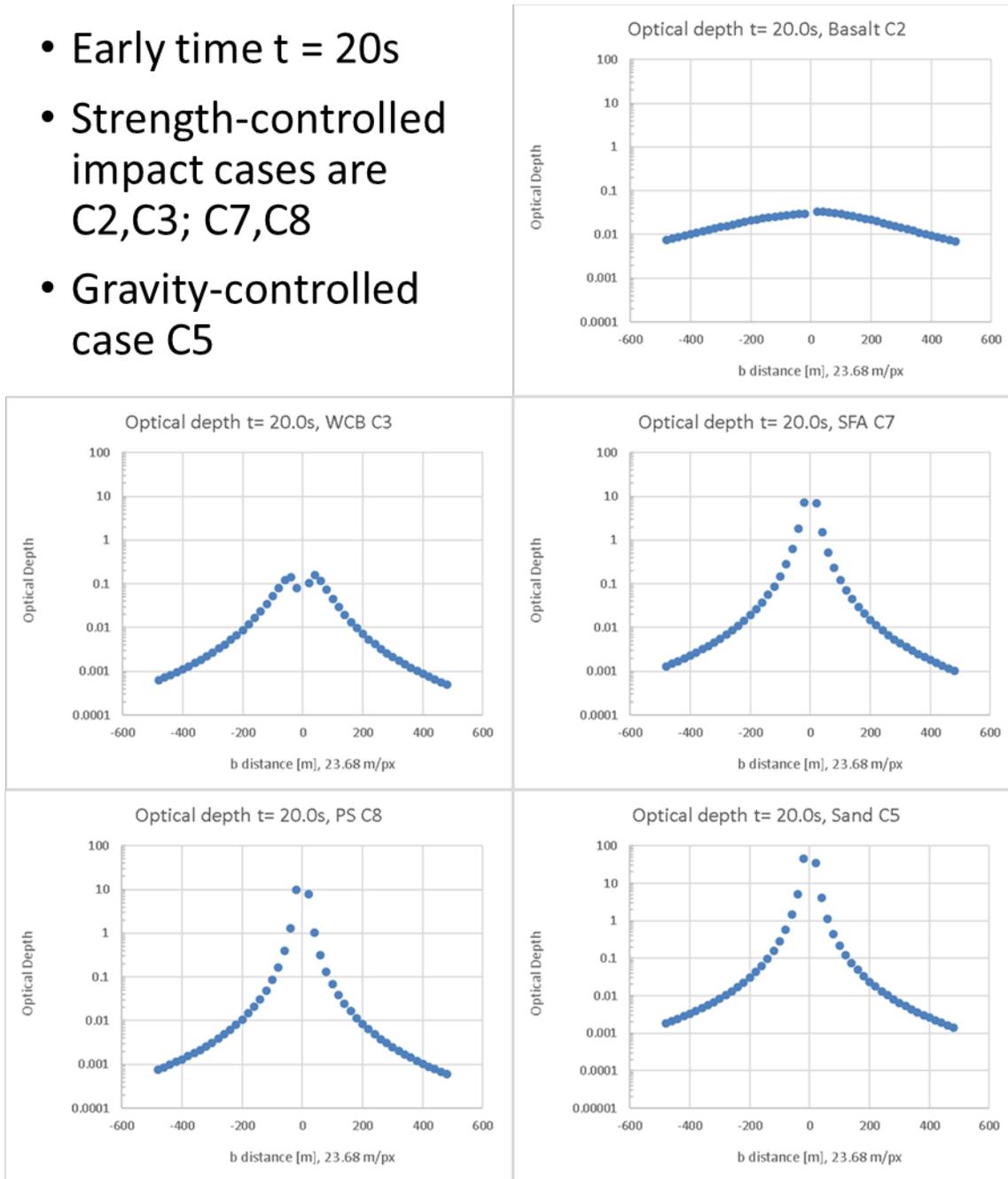

Figure 9. Same as Fig. 7, at time t=20 s after DART impact, where $\alpha + \varepsilon = 86.686°$. Compared to Figs. 7 and 8, clearing of plume optical depth continues for C2 and begins for C3. In weak target cases (bottom row), optical depth is increasing with time at all distances plotted.

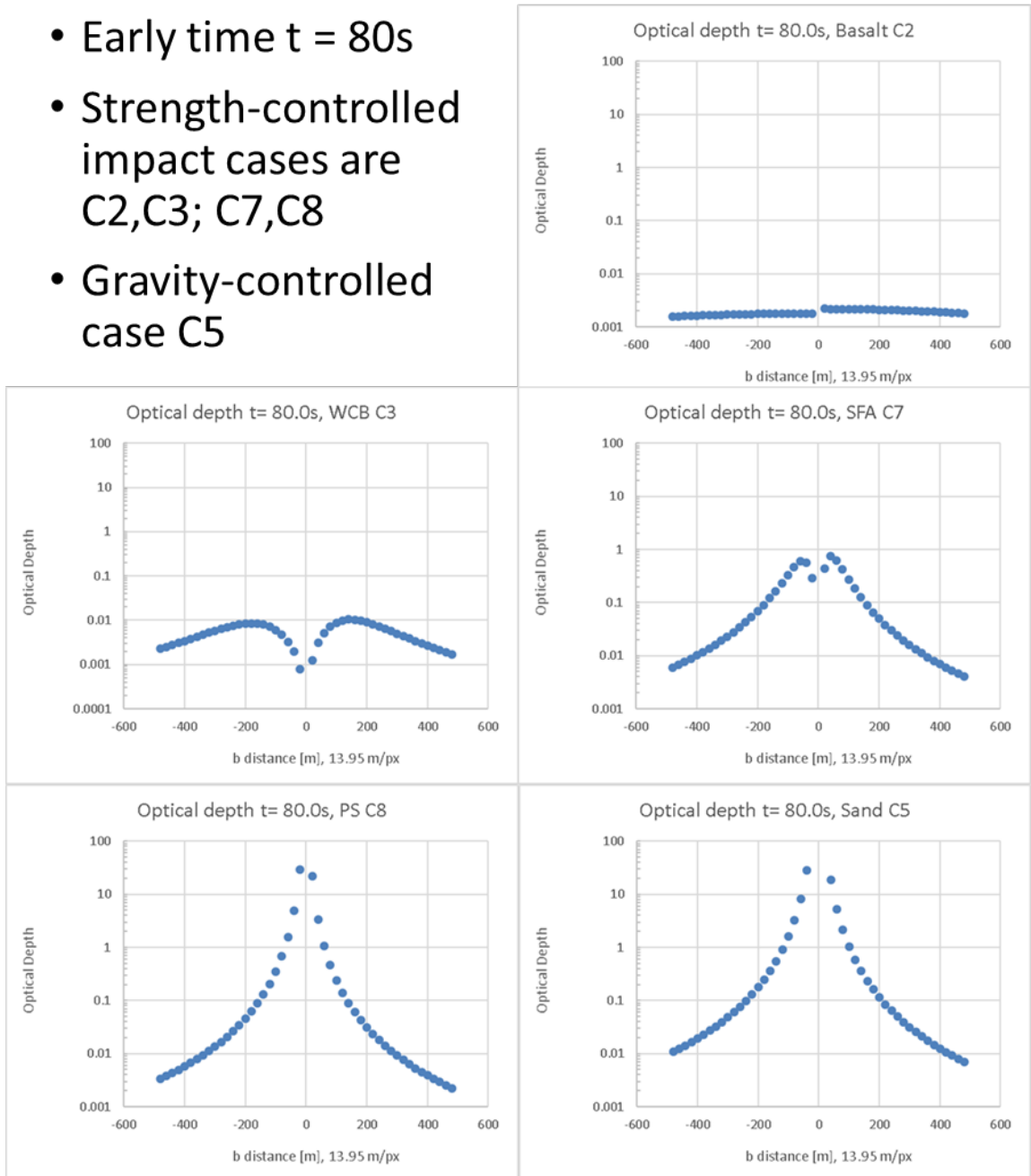

Figure 10 Same as Fig. 7, at time t=80 s after DART impact, where $\alpha + \varepsilon = 84.371°$. Continued clearing of cases C2, C3; clearing begins for case C7. In weakest target cases C8 and C5, optical depth is increasing with time at all distances plotted.

Fig. 10 shows that by *t*=80 s, for case WCB C3, the clearing of ejecta has proceeded out to |*b*| < 200 m, while optical depth is increasing with time at larger *b*. Also by *t*=80 s, clearing of the ejecta has started for case SFA C7, for low strength 4 kPa and 45% porosity. For

the still lower strength and higher porosity case C8 and the gravity-controlled case C5, there is still no clearing over the impact site.

Figs. 11 and 12 continue to show profiles at later times $t$=136 s and $t$=150 s, at which the PL1 resolution has reached 4.94 m/px and 2.86 m/px respectively. These profiles show an asymmetry between negative and positive $b$ values, as LICIACube approaches the ejecta cone crossing time when $x_s = y_s$; this time is t= 157 s. The asymmetry comes about because the LICIACube trajectory is $x_s = 55.4$ km and is offset from the asteroid (see Fig. 5). Although the plume is axisymmetric in the present model, it is not viewed along the symmetry axis, and observed profiles are asymmetric between positive and negative $b$.

The asymmetry noted in Figs. 11-13 arises for two reasons related to obliquity. Firstly, one side of the ejecta cone is viewed at higher obliquity (the side closer to LICIACube, at positive $b$) which increases optical path length. Secondly, the higher obliquity leads to viewing ejecta at higher altitude, implying higher velocity ejecta. Depending on the ejecta velocity distribution, the viewing of higher velocity ejecta may either increase or decrease the optical depth. In Figs. 11 and 12 for the case basalt C2, the two obliquity effects work together to yield a higher optical depth at positive $b$ versus negative $b$. In case C7, however, the two obliquity effects work together within $|b| < 100$ m but oppose one another otherwise. That is, the predominance of slower ejecta in the low strength case C7 is such that the optical depth at $b = -200$ m is an order of magnitude higher than at $b = 200$ m. Similar asymmetries are seen in the other weak target cases C8 and C5. Fig. 12 shows that clearing of the ejecta over the impact site has started for case PS C8.

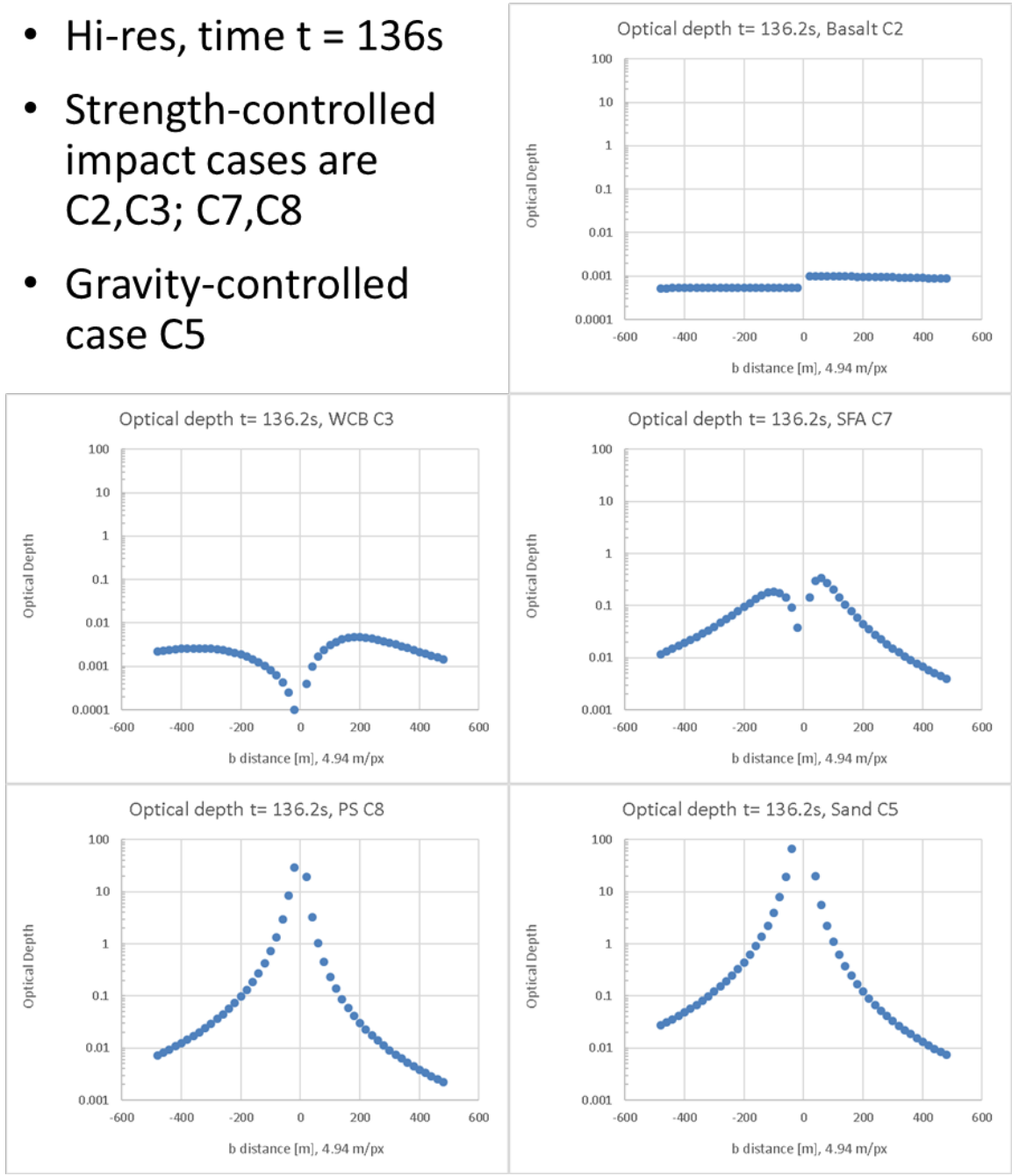

Figure 11. Same as Fig. 7 but at time t=136 s at which PL1 resolution is better than 5 m/px, where $\alpha + \varepsilon = 73.918°$. Asymmetry is noted between positive and negative $b$ from obliquity effects (see text).

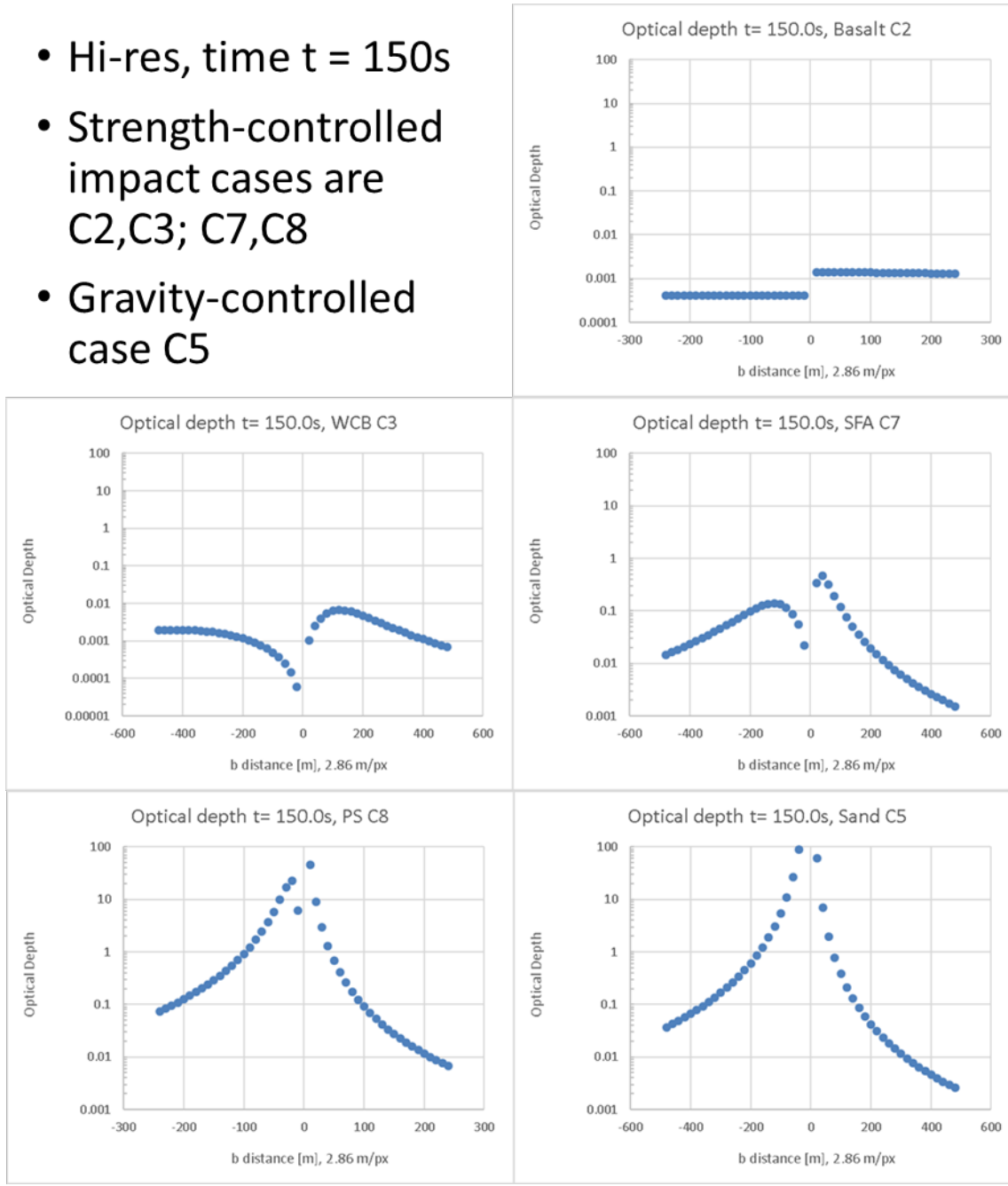

Figure 12. Same as Fig. 7 but at time t=150 s, where $\alpha + \varepsilon = 61.402°$. Increasing asymmetry from obliquity effects. Clearing begins for case C8.

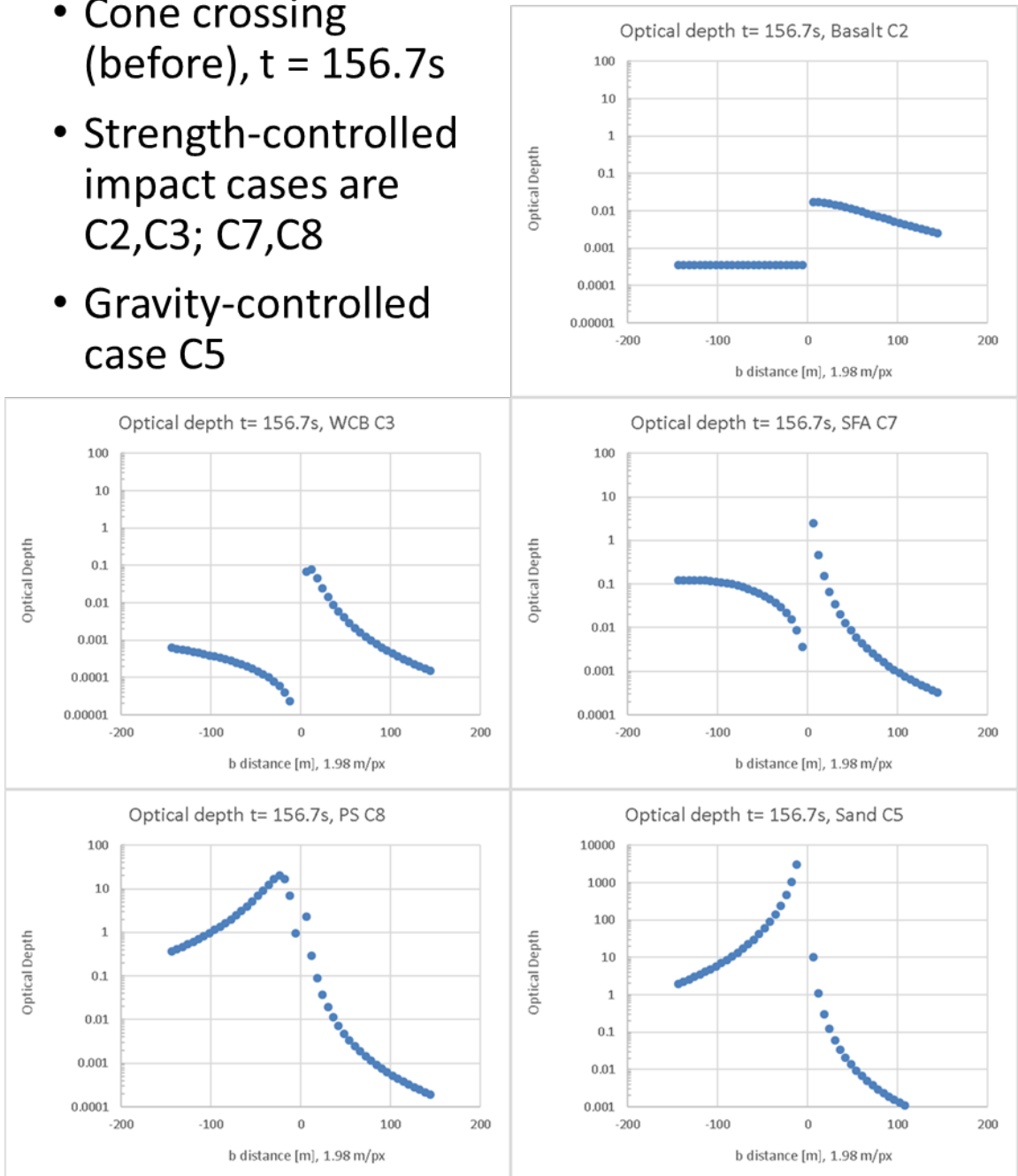

Figure 13. Same as Fig. 7 but at time t=156.7 s, where $\alpha + \varepsilon = 46.171°$. Close to time of ejecta cone crossing; maximal effect of obliquity enables estimation of ejecta cone thickness. Obliquity effect for case C5 reduces $\tau$ over one limb of target body, even though clearing has not yet begun.

Fig.13 shows an extreme asymmetry from the obliquity effect very near the time of cone crossing. This occurs in the present model because of the fixed ejection angle $\alpha = 45°$. A more realistic model would have the ejection angle varying over some range (Gulde et al. 2018) and thereby placing a limit on the maximum obliquity effect. LICIACube observations, by measuring the maximum asymmetry attained, can constrain the variation in ejection angle.

Also important to note is that in all five impact cases of the present model, including the gravity-controlled case where clearing does not start before LICIACube closest approach, at least one limb of the target asteroid Didymos B is visible through the plume at low optical depth. The limb of the target body behind the plume is at $|b| \approx 82$ m.

After the ejecta cone crossing and near closest approach to Didymos, LICIACube views the plume from outside the ejecta cone, with two intersections for each LOS (see Fig. 5) in which case the two optical depth contributions are summed. Fig. 14 at $t=164$ s shows an example of such optical depth profiles along the axis of the ejecta plume (although each LOS generally combines contributions from two altitudes). The profiles show the extent of clearing of the ejecta, from almost completed for case C2 to just having started for case C8; the distance $b$ at which the optical depth reaches a maximum is a useful discriminator between the impact cases. Clearing has not yet started for the gravity-controlled case C5, although the plume is optically thick at low altitudes (not plotted). The plume profiles obtained by LICIACube near closest approach, from outside the ejecta cone, can distinguish between all five of the impact cases. The LICIACube images from outside the ejecta cone also enable, importantly, direct measurements of the ejection angle $\alpha$.

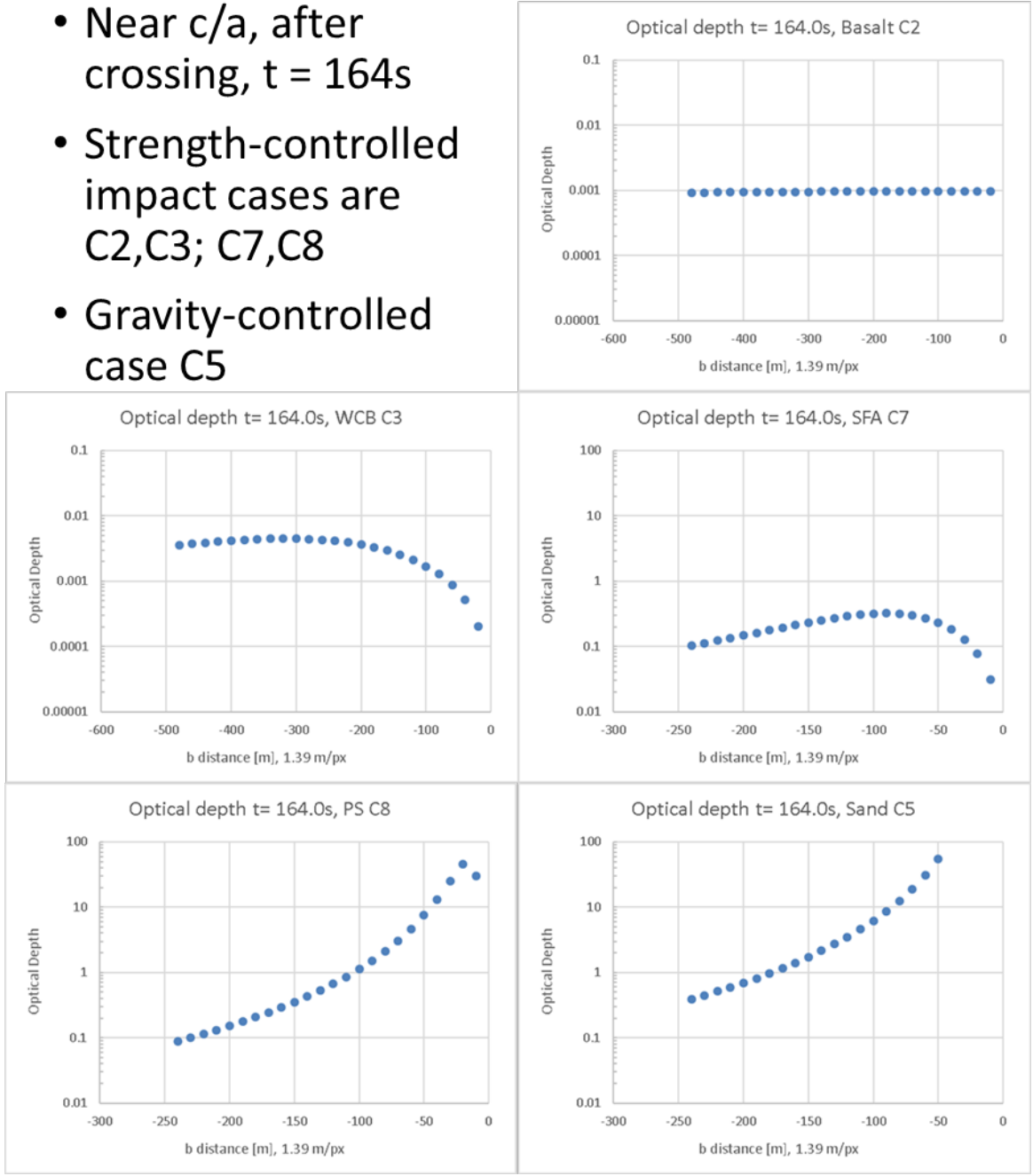

Figure 14. Same as Fig. 7 but at time t=164 s just before closest approach to Didymos, where $\alpha + \varepsilon = 9.769°$. Optical depth profiles along the axis of the plume, viewed from outside the ejecta cone with impact site at $b = 0$.

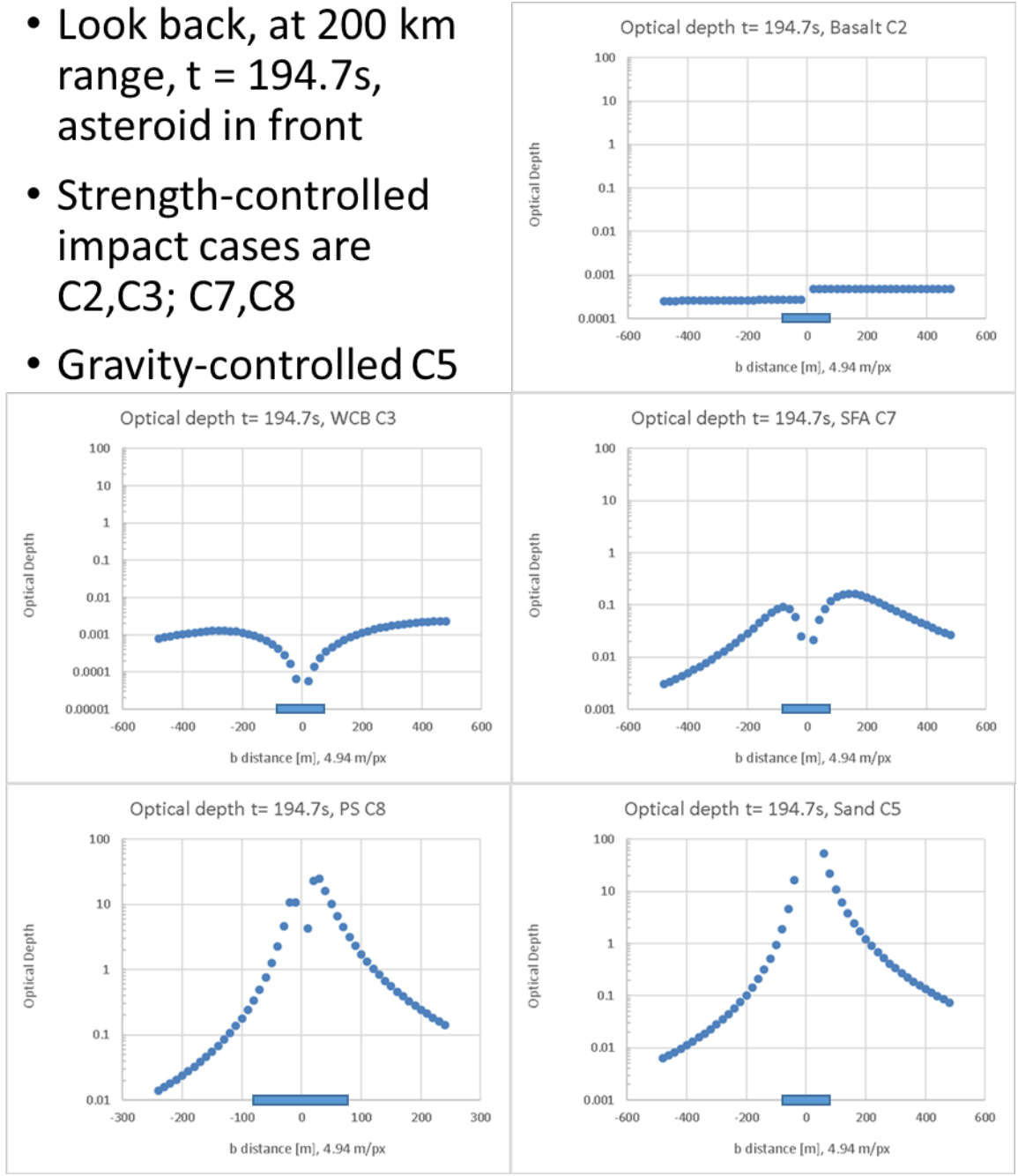

Figure 15. Same as Fig. 7 but at time t=194.7 s, after closest approach, where $\alpha + \varepsilon = -73.918°$. Plume is imaged above the limb of the target body, in forward scattering geometry at 120° phase angle. Region behind asteroid indicated by 160 m scale bars.

After closest approach, LICIACube continues its autonomous tracking of Didymos and slews to obtain images of the non-impact hemisphere and the ejecta plume above the limb of the target asteroid, which is now in front of the plume. Fig. 15 shows optical depth profiles at $t$=194.7 s, which are usefully compared to those in Fig. 11 at the same imager resolution. In Fig. 15 only the higher altitude portions of the profiles above the limb, $|b| >$ 82 m, are visible. Comparing to the higher altitude portions of the profiles in Fig. 11, for the gravity controlled case C5, the optical depth is ten times greater in Fig.15 than in Fig. 11 near $b \approx +200$ m but is four times less near $b \approx -200$ m. However, for the low strength case SFA C7, the optical depth shows lesser changes of the same sign, an increase by a factor 2-3 near $b \approx +200$ m but a similar decrease near $b \approx -200$ m. The optical depth changes for PS C8 are similar to those for the gravity-controlled case C5, with again an increase in optical depth near $b \approx +200$ m but a decrease near $b \approx -200$ m.

The effect of a different ejection angle on the optical depth profiles is illustrated in Fig. 16, which shows optical depth profiles calculated for $\alpha = 60°$ at the time $t$ =136 s, which is the time of Fig. 11. A change in ejection angle causes changes in the obliquity factor and in the LOS intersection with the ejecta cone (affecting the velocity sampled on the LOS). Importantly, it also changes the time of the ejecta cone crossing, which is 157.0 s at $\alpha = 45°$, but is 148.6 s at $\alpha = 60°$. At the time $t$ =150 s of Fig.12, LICIACube is already outside the cone for $\alpha = 60°$, whereas it is inside the cone for Fig. 12 with $\alpha = 45°$. Changes from Fig. 11 to Fig. 12 from temporal evolution are similar to those from different ejection angle, comparing Fig.11 to Fig. 16. However, the ejection angle is directly measureable from LICIACube images obtained outside the ejecta cone. It does not need to be inferred simultaneously with the inference of target material cases.

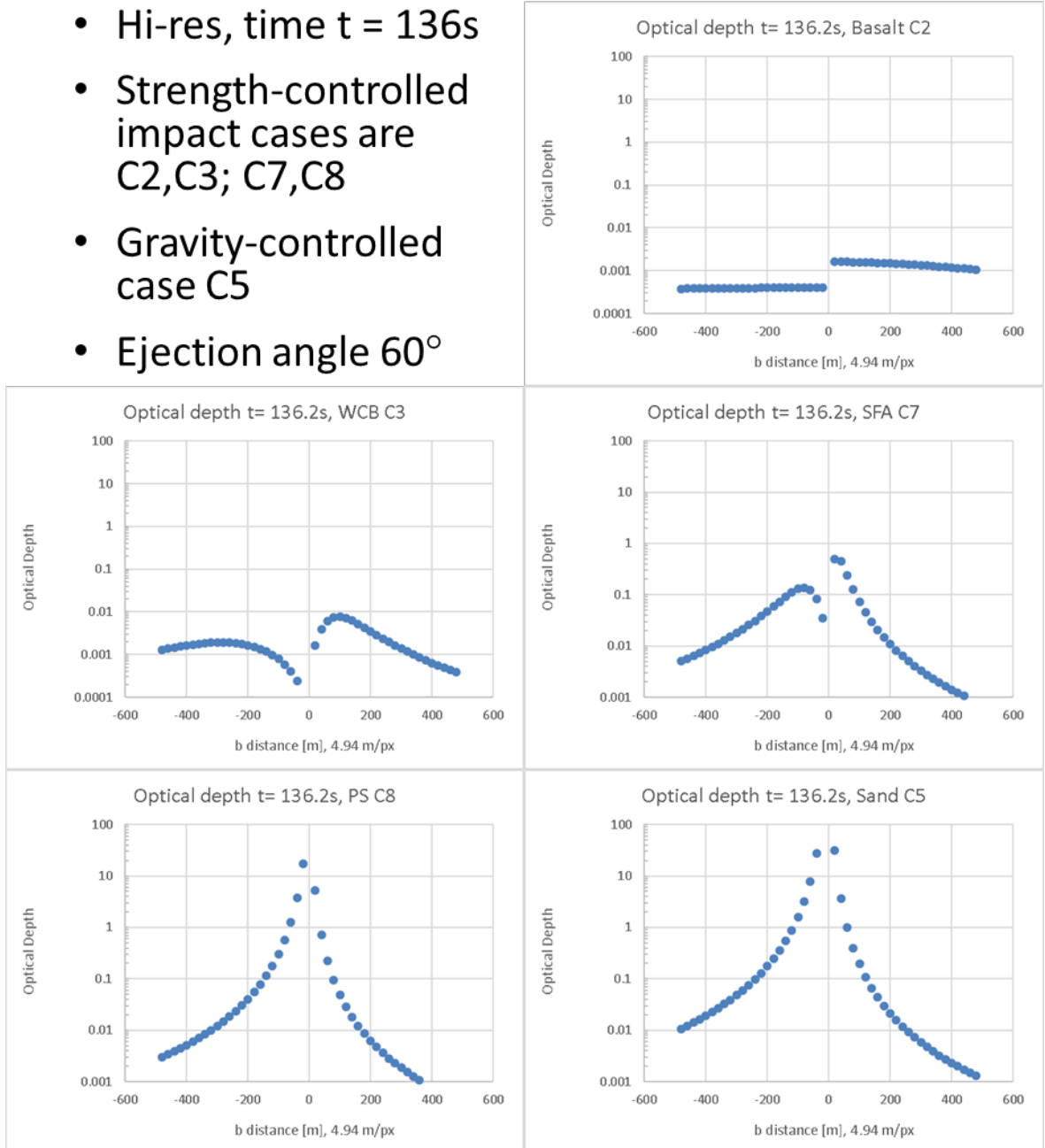

Figure 16. Same as Fig. 7, but at time t=136 s with different ejection angle 60°. Asymmetry effects are more pronouced than in Fig.11, which is at the same time t=136 s but with ejection angle 45°.

## 5. Discussion and Conclusions

The DART mission to demonstrate asteroid deflection by a kinetic impact will return fundamental new information on the responses of an asteroid to a hypervelocity impact.

DART will improve and validate models and simulations of kinetic impact to reduce the uncertainty of momentum transfer in future kinetic impactor missions, helping to validate the kinetic impactor technique and reducing risks for future asteroid hazard mitigation. The determination of the momentum transfer to Didymos B from the DART impact will provide the key new information to achieve the DART planetary defense objectives.

DART measurements of the Didymos orbital period change after the DART impact will determine the change in the transverse (circular) velocity component of Didymos B. With the mass of Didymos B, one component of the momentum transfer is determined. Information to determine the full vector momentum transfer is provided by: DART approach imaging that determines the DART impact site and characterizes its geology; numerical simulations of the DART impact that account for impact conditions as well as characteristics of the impact site and those of the spacecraft; and importantly, additional observations of Didymos B and of the DART impact ejecta plume by LICIACube during its separate flyby of Didymos after being released by DART. LICIACube will image the non-impact hemisphere of Didymos B, the side not visible to DART, and will image the ejecta plume structure and evolution.

The LICIACube plume observations provide important information that enhances the estimation of momentum transfer from the DART impact. The ejecta plume geometry can be determined from the LICIACube images to constrain the direction of the ejecta momentum $\boldsymbol{p}_{ej}$ by finding the direction of the plume axis and determining the asymmetry if any around this axis. In addition, the ejecta plume optical depth profiles and the changes with time after the DART impact enable characterization of the ejecta mass versus velocity distributions and then inference of target physical properties.

Figs. 7-15 illustrate how the plume profiles can distinguish all five of the impact cases considered in the present ejecta plume model. These cases include a gravity-controlled impact case and four strength-controlled impact cases with target properties ranging from very strong and nonporous to very weak and porous. The LICIACube plume observations will provide constraints on the target strength and porosity. Specific observables from the plume images include the time at which clearing of ejecta becomes evident over the impact site. This time is earlier for the stronger, less porous targets, as illustrated by the impact cases considered in Figs. 7-15: clearing times are ~10 s for basalt C2, ~20 s for WCB C3, ~80 s for SFA C7, ~150 s for PS C8; and for the gravity-controlled case, clearing does not begin before closest approach at 165.4 s. If clearing has started, the altitude at which the optical depth is at a maximum is another useful observable in profiles obtained near closest approach, when LICIACube images the plume along lines of sight almost perpendicular to the plume axis. The plume images after closest approach, when LICIACube has turned around to view the plume in forward scattering above the limb of the target body, provide additional information to discriminate between impact cases, particularly those at low strength and high porosity.

It is noted that LICIACube will obtain plume images from outside the ejecta cone near closest approach, viewing the cone from its side and allowing direct measurement of the ejection angle $\alpha$. The ejection angle does not need to be assumed or inferred from observations of plume structure and evolution simultaneously with target material properties. The discrimination between various target material cases can be made as illustrated in Figs. 7-15, but with an observed value of ejection angle $\alpha$.

It is emphasized that the observables from plume images that are used to discriminate impact cases and constrain target physical properties have not yet included the values of the optical depth, but instead have used the spatial and temporal variations of the optical depth. The actual values of the optical depth depend on the ejecta particle size distribution, for which the present model assumes an Itokawa-like, coarse regolith power law distribution with a minimum particle size of 1 mm. With this size distribution, in all five impact cases including the gravity-controlled case where clearing does not start before LICIACube closest approach, at least one limb of the target asteroid Didymos B is visible through the plume at low optical depth. This result also depends on the assumption in the present model of a fixed ejection angle which leads to a pronounced asymmetry from line-of-sight obliquity. The significance of observing the limb of the target body through the ejecta plume is that this observation enables determination of the plume optical depth as was done for Deep Impact at comet 9P/Tempel 1 (Kolokolova et al. 2016). If the plume optical depth can be determined, then the ejecta particle size distribution is determined while fitting optical depth profiles to the ejecta plume observations, better constraining target physical properties.

**Acknowledgements**

We thank NASA for support of the DART mission at JHU/APL under Contract # NNN06AA01C, Task Order # NNN15AA05T. Elisabetta Dotto and Vincenzo Della Corte acknowledge financial support from Agenzia Spaziale Italiana (ASI, contract No. 2019-31-HH.0 CUP F84I19001260O). We thank Thomas Statler and Steven Chesley for helpful discussions of the momentum transfer efficiency β.

**References:**

A'Hearn, M. F. et al. 2005. Deep Impact: Excavating Comet Tempel 1. Science, 310, 258

Arakawa, M. et al. 2020. Science 10.1126/science.aaz1701

Biele, J., Ulamec, S., Malbaum, M., Roll, R., Witte, L. et al. 2015. The landing(s) of Philae and inferences about comet surface mechanical properties. Science, 349, doi: 10.1126/science.aaa9816

Bruck Syal, M., Owen, J.M., Miller, P.L. 2016. Deflection by kinetic impact: sensitivity to asteroid properties. Icarus, 269, 50-61

Buhl, E., Sommer, F., Poelchau, M., Dresen, G., Kenkmann, T. 2014. Ejecta from experimental impact craters: Particle size distribution and fragmentation energy. Icarus, 237, 131-42

Cheng, A. F. and Hayabusa Team. 2009. Fundamentally Distinct Outcomes of Asteroid Collisional Evolution. Planetary Sp. Sci., 57, 165-172. doi:10.1016/j.pss.2008.08.013

Cheng A. F. et al. 2016. Asteroid Impact & Deflection Assessment mission: Kinetic Impactor. Planet. Space Sci., 121, 27–35

Cheng A. F. et al. 2018. AIDA DART asteroid deflection test: Planetary defense and science objectives. Planet. Space Sci., 157, 104-115

Colaprete, A. et al. 2010. Detection of Water in the LCROSS Ejecta Plume. Science, 330, 463-468

Feldhacker, J.D., Bruck Syal, M., Jones, B.A., Doostan, A., McMahon, J., Scheeres, D. 2017. Shape Dependence of the Kinetic Deflection of Asteroids. J.Guidance Control Dyn. 40, 2417-2431

Fu, Q. and Sun, W. 2001. Mie theory for light scattering by a spherical particle in an absorbing medium. Applied Optics, 40, 1354-61.

Gulde, M., Kortmann, L., Ebert, M., Watson, E., Wilk, J., Shaefer, F. 2018. Robust optical tracking of individual ejecta particles in hypervelocity impact experiments, MAPS, 53, 1696-1704

Hermalyn, B. et al. 2012. Scouring the surface: Ejecta dynamics and the LCROSS impact event. Icarus, 218, 654-665

Holsapple K. and Housen K. 2012. Momentum transfer in asteroid impacts. I. theory and scaling. Icarus, 221, 875-887

Housen, K. and Holsapple,K.,2011. Ejecta from impact craters. Icarus, 211,856–875.

J. M. Trigo-Rodríguez, J. Llorca. 2007. The strength of cometary meteoroids: Clues to the structure and evolution of comets. Mon. Not. R. Astron. Soc. 372, 655–660 2006; erratum: Mon. Not. R. Astron. Soc. 375, 415–415

Jutzi, M.and Michel,P.,2014.Hypervelocity impacts on asteroids and momentum transfer. I. Numerical simulations using porous targets. Icarus 229: 247–253.

Kolokolova, L., Nagdimunov, L., A'Hearn, M., King, A., Wolff, M. 2016. Studying the nucleus of comet 9P/Tempel 1 using the structure of the Deep Impact ejecta cloud at the early stages of its development. PSS, 133:76-84.

Luther, R., Zhu, M-H, Collins, G., Wuennemann, K. 2018. Effect of target properties and impact velocity on ejection dynamics and ejecta deposition. MAPS, 53, 1705-1732.

Michel P. et al. 2016. Science case for the Asteroid impact Mission (AIM): a component of the Asteroid Impact & Deflection Assessment (AIDA) mission. Adv. Space Res., *57*, 2529-2547

Michel, P. et al. 2018. European component of the AIDA mission to a binary asteroid: Characterization and interpretation of the impact of the DART mission. Adv. Space Res., 62, 2261-72.

Naidu, S., et al. 2020. Radar observations and a physical model of binary near-earth asteroid 65803 Didymos, target of the DART Mission. Icarus, 348, 113777, https://doi.org/10.1016/j.icarus.2020.113777.

Nakamura E., Makishima A., Moriguti T., Kobayashi K., et al. 2012. Space environment of an asteroid preserved on micrograins returned by the Hayabusa spacecraft. PNAS, doi:10.1073/pnas.1116236109

O'Keefe, J. and Ahrens T. 1993. Planetary Cratering Mechanics. JGR, 98: 17011-17028

Owen, J.M., Bruck Syal, M., Remington, T., Miller, P.L., Richardson, D., Asphaug, E., 2017. Modeling kinetic impactors on a rubble pile asteroid. In: IAA Planetary Defense Conference. IAA-PDC-17-04-08, Tokyo, Japan (abstract).

Prieur, N. C., Rolf, R., Luther, R., Wuennemann, K., Xiao, Z., Werner, S.c. 2017. The effect of target properties on transient crater scaling for simple craters. JGR, 122, 1704-1726

Raducan, S.; Davison, T.M.; Luther, R.; Collins, G.S.; The Role of Asteroid Strength, Porosity and Internal Friction in Impact Momentum Transfer. 2019. Icarus, 329, 282-295.

Scheirich P. and Pravec P. 2009. Modeling of light curves of binary asteroids. Icarus, 200, 531.

Schultz, P.H. et al. 2010. The LCROSS Cratering Experiment. Science, 330, 468-472

Stickle, A.M. Atchison, J., Barnouin, O., Cheng, A.F., Crawford, D., Ernst, C., Fletcher, Z., Rivkin, A. 2015. Modeling Momentum Transfer from Kinetic Impacts: Implications for Redirecting Asteroids. Procedia Eng. 103, 577-584.

Stickle, A.M., Rainey, E., Owen, J.M., Raducan, S., Bruck Syal, M. et al. 2018. Modeling Momentum Enhancement from Impacts into Rubble Pile Asteroids. LPSC 49, LPI 2083, id.1576

Stickle, A.M., Bruck Syal, M., Cheng, A. F., Collins G. S., et al. 2020. Benchmarking impact hydrocodes in the strength regime: Implications for modeling deflection by a kinetic impactor. Icarus, 338,113446

Thomas, N., Sierks, H., Barbaieri, C., Lamy P. et al. 2015. The morphological diversity of comet 67P/Churyumov-Gerasimenko. Science, 347; DOI: 10.1126/science.aaa0440